\documentclass[
    ,final            
  ]
  {aipproc}

\layoutstyle{6x9}
\def\Chandra{${\it Chandra}$\ }
\def\HST{${\it HST}$\ }
\newcommand{\Msun}{\ifmmode {M_{\odot}}\else${M_{\odot}}$\fi}
\newcommand{\Lsun}{\ifmmode {L_{\odot}}\else${L_{\odot}}$\fi}

\begin{document}

\title{X-ray Binaries in the Globular Cluster 47 Tucanae}


\author{C~O.~Heinke}{
  address={Northwestern University, 
2131 Tech Drive,
Evanston, IL 60208}
  ,altaddress={Harvard University, 
60 Garden Street, 
Cambridge, MA 02138} 
}

\author{J.~E.~Grindlay}{
  address={Harvard University, 
60 Garden Street, 
Cambridge, MA 02138}
}

\author{P.~D.~Edmonds}{
  address={Harvard University, 
60 Garden Street, 
Cambridge, MA 02138}
}

\author{H.~N.~Cohn}{
  address={Indiana University, 
Dept. of Astronomy, Swain West 319, 
Bloomington, IN 47405}
}

\author{P.~M.~Lugger}{
  address={Indiana University, 
Dept. of Astronomy, Swain West 319, 
Bloomington, IN 47405}
}
\author{F.~Camilo}{
  address={Columbia Astrophysics Laboratory,
 550 West 120th Street,
New York, NY 10027}
}
\author{S.~Bogdanov}{
  address={Harvard University, 
60 Garden Street, 
Cambridge, MA 02138}
}
\author{P.~C.~Freire}{
  address={NAIC, Arecibo Observatory,
HC03 Box 53995,
PR 00612}
}

\begin{abstract}
\Chandra observations of globular clusters provide insight into the
formation, evolution, and X-ray emission mechanisms of X-ray
binary populations.  Our recent (2002) deep observations of 47 Tuc allow
detailed study of its populations of quiescent LMXBs, CVs, MSPs, and
active binaries (ABs). 
First results include the confirmation of a magnetic CV in a globular
cluster, the identification of 31 additional chromospherically active
binaries, and the identification of three additional likely quiescent LMXBs
containing neutron stars.  Comparison of the X-ray properties of the
known MSPs in 47 Tuc with the properties of the sources of
uncertain nature indicates that relatively few X-ray sources are MSPs,
probably only $\sim$30 and not more than 60.  Considering the $\sim$30
implied MSPs and 5 (candidate) quiescent LMXBs, and their canonical
lifetimes of 10 and 1 Gyr respectively, the relative birthrates of
MSPs and LMXBs in 47 Tuc are comparable. 
\end{abstract}

\maketitle


\section{Introduction}
Globular clusters are efficient factories for the production of X-ray
binaries from populations of primordial binaries (see papers by
D'Antona et al. and Ivanova et al., this volume).  Although the nature
of bright ($L_X>10^{36}$ ergs s$^{-1}$) X-ray sources in globular clusters as
low-mass X-ray binaries containing neutron stars has been long
established \cite{Lewin83, Grindlay84}, the nature of faint X-ray
sources has been more difficult to determine.  At least four types of
low-luminosity ($L_X=10^{30-34}$ ergs s$^{-1}$) X-ray systems are now 
known to exist in globular clusters:  cataclysmic variables
\cite[CVs;][]{Hertz83}, LMXBs in quiescence
\cite[qLMXBs;][]{Hertz83,Verbunt84}, millisecond radio pulsars
\cite[MSPs;][]{Saito97}, and chromospherically active main-sequence
binaries \cite[ABs;][]{Bailyn90b}.  A recent review of X-ray sources
in globular clusters can be found in \citet[][see also Verbunt, these
  proceedings]{Verbunt04}. 

The advent of the \Chandra X-ray Observatory has allowed detailed
study of the populations of faint X-ray sources in globular clusters 
\cite{Pooley03},  
especially in the dense and nearby cluster 47 Tuc \cite{Grindlay01a}.
The large MSP population in 47 Tuc \cite{Camilo00} has been detected
in X-rays with {\it Chandra}, allowing studies of the MSPs' X-ray
luminosities and spectra \cite{Grindlay02,Bogdanov04}.  The precise
positions have allowed identification of optical counterparts to 58\%
(45 of 77) of the X-ray sources detected within a deep \HST imaging
field \cite{Edmonds03a,Edmonds03b}.  Within 47 Tuc, 22 CVs and
29 ABs have been unambiguously identified, in addition to two 
qLMXBs and 17 MSPs.

A major mystery associated with globular clusters is the formation
mechanism of millisecond pulsars.  The logical progenitors of MSPs,
the LMXBs, appear at first glance to be far too small in numbers to
produce the estimated numbers of MSPs \cite{Kulkarni90}.  The number
of MSPs in 47 Tuc 
has been estimated at $>$200 \cite{Camilo00}, suggesting $\sim$10000 MSPs
in the Galactic globular cluster system.  However, only 13 bright
LMXBs are known in our globular cluster system.  For typically assumed
lifetimes of 10 and 1 Gyr respectively, the MSP birthrate is two
orders of magnitude higher than the LMXB birthrate.  It is possible
that many MSPs in globular clusters were born in the distant past from
intermediate-mass 
X-ray binaries, and/or that X-ray heating greatly reduces the
lifetimes of LMXBs \cite{Pfahl03}.  However, the problem cannot be
regarded as conclusively solved via these theoretical mechanisms at
the present time.  Here we discuss observations of 47 Tuc which may
 reduce this discrepancy in lifetimes by two orders of magnitude.

\section{\Chandra Observations of 47 Tuc}
Following our successful 70 ksec \Chandra observation of 47 Tuc in 2000
\cite{Grindlay01a}, we obtained a deeper 280 ksec observation in late
2002, spread over more than a week to constrain source variability.  
These observations used the backside-illuminated ACIS-S3 chip for
maximal low-energy sensitivity.  Within the 2.79$'$ half-mass radius
of 47 Tuc, 300 sources were detected in the 2002 observations
\cite[Figure 1; Grindlay, these proceedings;][]{Heinke04a}. 

\begin{figure}
  \includegraphics[height=.6\textheight]{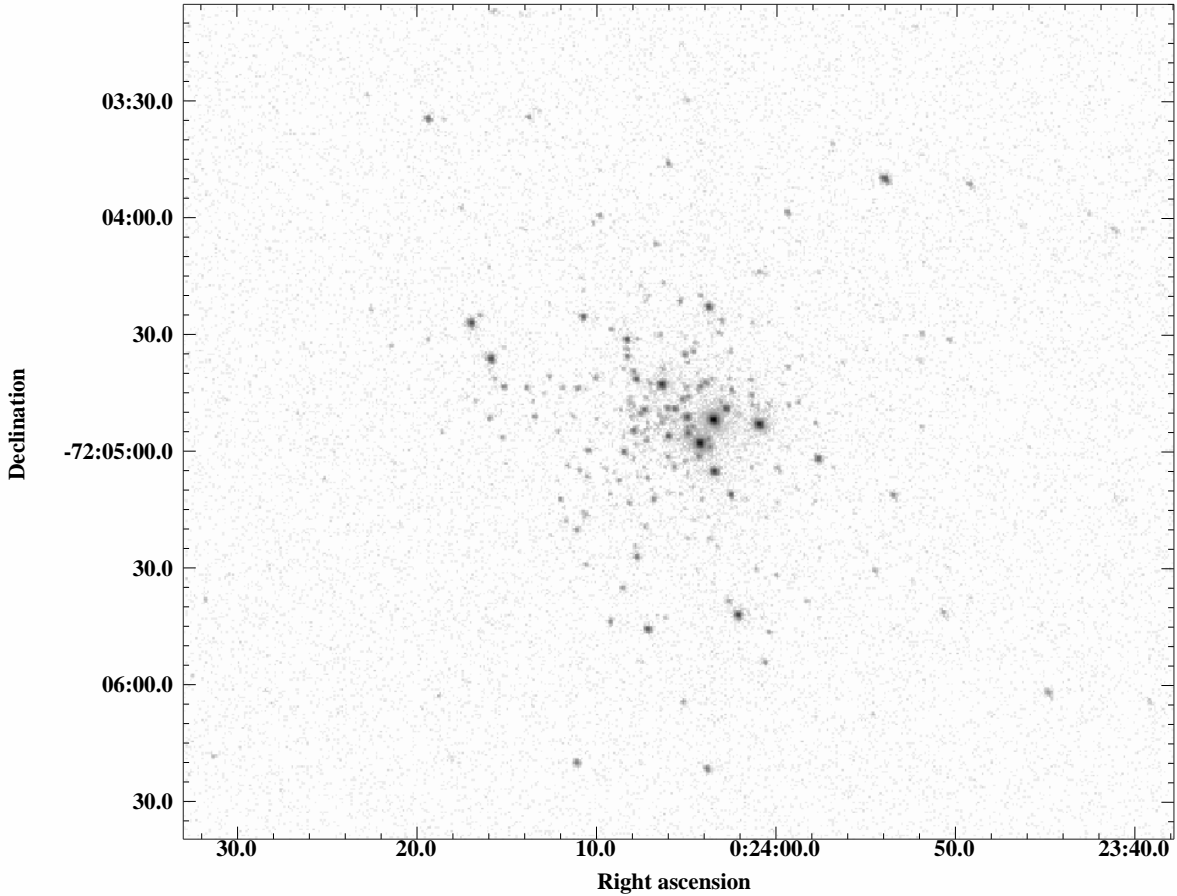}
  \caption{\Chandra X-ray image of 47 Tuc, 0.3-6 keV energy range. The
  corners of this view are located at the half-mass radius.}
\end{figure}\label{fig:box}

Comparing the positions of the new X-ray sources with unusual objects
 in 47 Tuc, we find that an additional 31 active binaries, 
identified from \HST variability studies
 \cite{Albrow01}, can be unequivocally identified with \Chandra X-ray
 sources. None of the new CV candidates suggested by \cite{Knigge02}
 are identified with X-ray sources.   Eighty-seven X-ray sources
show clear variability, on timescales ranging from hours and weeks
(within the 2002 observations) to years (comparing the 2002 and 2000
observations) and decades (comparing the 2002 detections with the
ROSAT results, \cite{Verbunt98}).  Based on the radial distribution of
 the detected sources, roughly 70 of them are probably background
 sources.  

Simple spectral fits to the individual sources find that the majority
can be well-fit by absorbed thermal plasma models (VMEKAL in XSPEC)
using the cluster metallicity.  This is expected for CVs and ABs
(though some of the brightest CVs require complicated multi-component
spectra).  However, the known qLMXBs and MSPs are generally not
well-fit by thermal plasma models \cite{Heinke04a, Bogdanov04}.  There
are three additional sources that are unusually soft and bright when
compared to other CVs and ABs.  

One, W37, shows eclipses with a 3.087
hour period and strong variations in $N_H$, with an X-ray spectrum
generally best
described by a hydrogen-atmosphere model appropriate for neutron star
surfaces \cite{Lloyd03}.  The other two, X4 and W17, have X-ray
spectra best described by a hydrogen-atmosphere model with a strong
power-law component.  X4 also shows short-term variability, which may
be from either component.  The hydrogen-atmosphere models give implied
radii consistent with 10-13 km for all three sources, indicating that
all three are probably quiescent LMXBs.  However, the relative
strength of the power-law components (60-65\% of the 0.5-10 keV
unabsorbed flux) and low luminosities ($\sim5\times10^{31}$ ergs
s$^{-1}$) for two, and high absorption for the third, indicate that
qLMXBs such as these would probably not have been identified in other
globular clusters \cite{Heinke04b}.  

Cataclysmic variables in globular clusters seem to have higher X-ray 
fluxes, and lower optical fluxes, than CVs in the rest of the Galaxy 
\cite{Edmonds03b}.  This has led to speculation that the accretion flow 
onto CVs in globular clusters may be generally controlled by the 
magnetic field of the white dwarf, since such systems (DQ Her and 
AM Her systems) have little or no 
disk and relatively high X-ray/optical flux ratios \cite{Grindlay96}.  
However, only one DQ Her-type system has been suggested in a globular 
cluster \cite[X9 in 47 Tuc,][]{Grindlay01a}.  Our new \Chandra 
observations identify a large-amplitude sinusoidal modulation with a 
4.7-hour period from the CV X10, and a very soft blackbody component
to its X-ray spectrum, identifying this CV as an AM Her, or polar.  
We also identify an $N_H$ column above the cluster value for 12 of 
the 22 CVs, possibly indicating a DQ Her nature.  Period searches are 
underway to confirm this possibility (Grindlay et al. 2005, in prep.).

\section{Millisecond Pulsars in 47 Tuc}

Of the 22 known MSPs in 47 Tuc, 17 have known positions, 16 through
radio timing \cite{Camilo00,Freire01a}, and one through matching an X-ray
source to a variable optical source with an orbital period and phase that
match those of 47 Tuc-W \cite{Edmonds02b}.  The MSPs G and I,
separated by only 0.1$''$, are detected as a single source in the \Chandra
image.  The MSPs F and S, separated by only 0.74$''$, are detected as a
single, extended source.  All other MSPs
with known positions are clearly detected in this image. 

Comparing the independently detected MSPs' radio pseudoluminosities
\cite{Camilo00} with their X-ray luminosities, no correlation is seen.
This implies that MSPs with lower radio pseudoluminosities will have
X-ray luminosities similar to those of the known MSPs.  
The X-ray emission is probably mostly from the hot polar caps of the
neutron stars \cite{Grindlay02}, while the radio emission
originates higher in the magnetosphere.  Gravitational bending assures
that we will see virtually all MSPs in the X-ray, regardless of radio
beaming fractions \cite{Bogdanov04}.  Thus, nearly all MSPs in 47 Tuc
should be detected among our X-ray sources, with X-ray luminosities 
between $2\times10^{30}$ and $2\times10^{31}$ (30-350 counts).

We can compare the X-ray properties (X-ray ``colors'', variability,
and spectral fits) of the unknown sources in this luminosity range 
to those of the known MSPs, 
CVs and ABs.  We show in Figure 2 the distributions of MSPs, good
candidate MSPs, ABs, and unknown sources in an X-ray color-color
plot. (Likely background sources, located farther than 100$''$ from the
center of 47 Tuc, are also indicated, and are somewhat harder on average.) 
The distributions of MSPs and ABs are seen to be significantly
different, and the unknown source distribution is more similar to the
ABs than the MSPs.

\begin{figure}
  \includegraphics[height=.4\textheight]{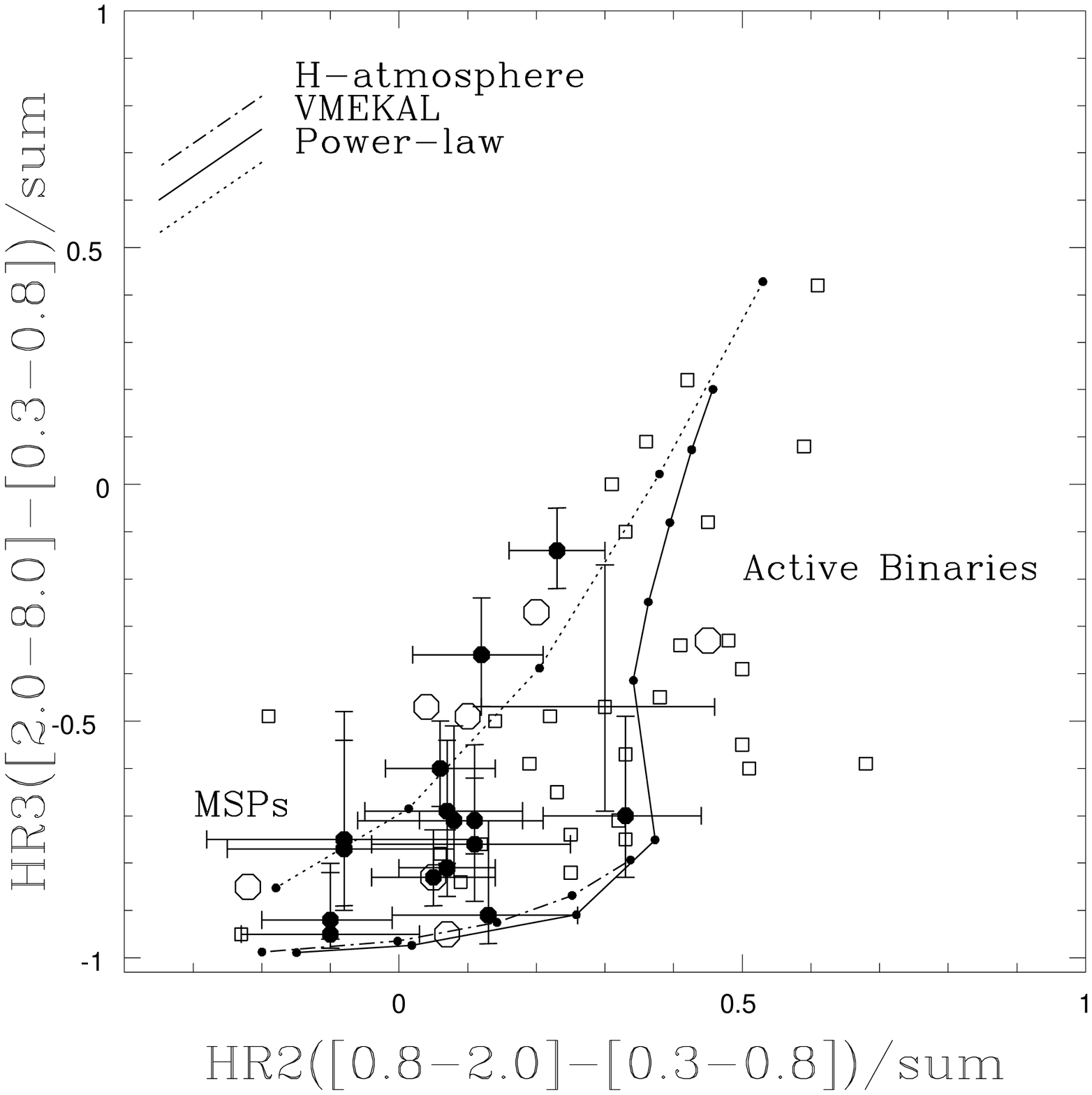}
  \includegraphics[height=.4\textheight]{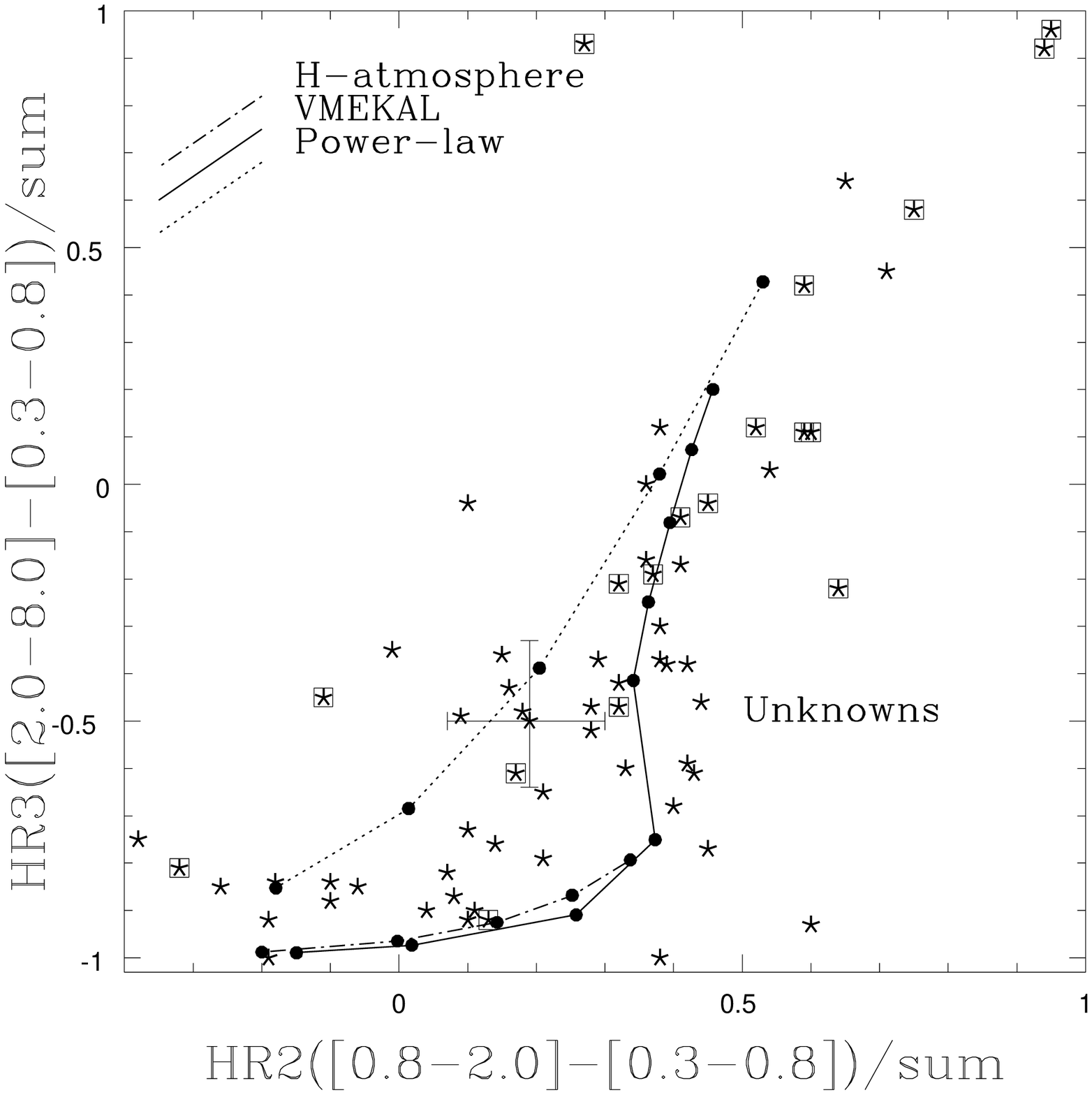}
  \caption{X-ray color-color plots showing sources with 30-350 counts
  in 47 Tuc.  Left: Locations of MSPs (large filled dots), candidate MSPs
  (open octagons), and ABs (open squares).  Right: Locations of
  sources of unknown nature (stars), and the subset of unknown sources
  located beyond 100$''$ from the center of the cluster (stars enclosed
  by squares); the latter are likely dominated by background sources.
  Model spectral tracks are plotted in both, with small dots
  representing (lower left to upper right): H-atmosphere for 75, 100,
  125, 150, 175 eV; VMEKAL thermal plasma for 0.4, 0.5, 0.7, 1, 2, 3,
  5, 10 and 30 keV; and a power-law with photon index 3, 2.5, 2, 1.5,
  1. Error bars are plotted for all MSPs, and a few other representative
  faint sources.  Increasing $N_H$ moves sources up and to the right.}
\end{figure}\label{fig:colorcolor}

We compare the numbers of MSPs, ABs, CVs, and background sources with
each property (variability, spectral agreement with a VMEKAL model,
colors within one of three classes) to the total number of unknown
sources with that property.  We assume that the ratio of unknown
members of a class (e.g. ABs) to the known members of that class is
the same (within binomial statistics) for each property.  That is, if
the number of ABs not yet identified is 60\% of the number of
identified ABs, and 20 known ABs are variable, then 
$\sim$12 unknown variable sources are probably ABs.  We fit the 
numbers of unknown sources with each property using the numbers of
known sources, to determine the ratios of unknown to known sources in
each class.  We can thus estimate the 
number of unidentified MSPs as $7^{+10}_{-7}$.  These unidentified MSPs 
must include the five known MSPs without known positions, suggesting 
that few MSPs remain undetected.  The 95\% single-sided upper limit 
on the total number of MSPs is 42; varying our choice of information to 
include gives a range of upper limits up to 56.  Because our CV sample is 
X-ray selected, we cannot provide a similar constraint upon the
 number of CVs.

Can this estimate be missing a significant number of radio MSPs?  
Radio-faint MSPs could be X-ray faint also.  However, there is no 
evidence of such a correlation among the known MSPs.  Submillisecond 
pulsars or pulsars in very tight orbits could escape radio detection, 
but since these would be relatively energetic they would 
tend to be X-ray brighter, not fainter, than the known MSPs.  The most 
likely way for MSPs to remain hidden is to be completely enshrouded in 
ionized gas from a nondegenerate companion (see Grindlay, these 
proceedings), similar to (but more extreme 
than) 47 Tuc W \cite{Camilo00,Bogdanov04}.  This would obscure the 
radio signal, and the shock from the pulsar would also emit X-rays, 
altering the X-ray emission spectrum.  However, 47 Tuc W is rather 
luminous for an MSP (and since the X-rays from the shock would need 
to obscure the thermal component, any ``hidden'' MSPs would be 
similarly luminous).  There are very few X-ray sources of unknown 
nature in 47 Tuc of similar luminosity as 47 Tuc W.
We conclude that unless there exists a class of MSPs that are very 
different from the MSPs we know, there are probably not more than 60 
MSPs in 47 Tuc.  This estimate agrees with independent constraints 
from \HST identifications of X-ray sources \cite{Edmonds03b} and 
integrated radio continuum flux measurements \cite{McConnell04}.

We extrapolate from this constraint in two directions.  Two lines of 
evidence suggest that of order 10\% of the neutron stars in a dense 
cluster are recycled into MSPs.  Six of the eighty known MSPs in 
globular clusters have companions of masses $\sim0.1-0.2$ \Msun, yet 
are eclipsing.  All of these systems with optical identifications are
probably MSP--main sequence binaries, suggesting binary exchange of a
white dwarf for a main sequence star and putting the pulsar on a path
to further recycling.  
Since 7\% of these pulsars were doubly recycled, the recycling rate for 
neutron stars in dense clusters is likely to be of the same order.
 Recent cluster simulations \cite{Ivanova03} 
also suggest a neutron star recycling rate of 5-15\%, higher for denser 
clusters.  Thus we can extrapolate a total number of $\sim300$ neutron 
stars in 47 Tuc (a rough, but empirical estimate).  This is 
significantly less than has been previously 
predicted, and helps to resolve the neutron star retention problem in 
globular clusters \cite{Pfahl02a}.

The other direction is toward relative LMXB/MSP birthrates.  These 
studies of 47 Tuc have increased the inferred number of qLMXBs and 
decreased the inferred number of MSPs, producing a ratio of 
$\sim$30/5=6 (compare to a ratio of $\sim$1000 from \cite{Kulkarni90}).  If 
qLMXBs can be counted among the progenitors of MSPs as 
a stage in the outburst cycle of LMXBs, and LMXB/qLMXBs live for 
$\sim$1 Gyr, then there is no longer an MSP birthrate problem 
in 47 Tuc, and in globular clusters generally.


\begin{theacknowledgments}
 We thank the Penn State team for the ACIS\_EXTRACT software, and D. Lloyd 
for his neutron star atmosphere models.  C.~H. was supported in part by 
\Chandra grant GO2-3059A, and by the Lindheimer fund at Northwestern 
University.
\end{theacknowledgments}



\bibliographystyle{aipproc}   

\bibliography{src_ref_list}

\IfFileExists{\jobname.bbl}{}
 {\typeout{}
  \typeout{******************************************}
  \typeout{** Please run "bibtex \jobname" to obtain}
  \typeout{** the bibliography and then re-run LaTeX}
  \typeout{** twice to fix the references!}
  \typeout{******************************************}
  \typeout{}
 }

\end{document}